\documentclass{phbauth}        
\usepackage{graphicx}

\begin{document}

\begin{frontmatter}

\title{Dynamic scaling of $I$-$V$ data for 
the neutral 2D Coulomb gas}

\author[address1]{Stephen W. Pierson\thanksref{thank1}},
\author[address2]{Mark Friesen}

\address[address1]{Department of Physics, Worcester Polytechnic Institute,
Worcester, MA 01609-2280, USA} \address[address2]{Applied
Superconductivity Center, University of Wisconsin, Madison, Wisconsin
53706, USA}

\thanks[thank1]{Corresponding author. E-mail: pierson@wpi.edu; FAX: (508)
831-5391} 

\begin{abstract}
The value of the dynamic critical exponent $z$ has been studied for 
experimental two-dimensional superconducting and Josephson
Junction array systems in zero magnetic field via the Fisher-Fisher-Huse
dynamic scaling analysis.  We found $z\simeq5.6\pm0.3$, a relatively large
value indicative of non-diffusive dynamics. We extend this work here to
simulational $I$-$V$ curves that are also found to be characterized by the
same large value of $z$. \end{abstract}

\begin{keyword}
Kosterlitz-Thouless dynamics; Dynamic Scaling; I-V characteristics
\end{keyword}

\end{frontmatter}

Dynamical scaling is a powerful tool for analyzing the dynamic 
critical behavior\cite{hohenberg} of a variety of systems. Fisher, 
%critical behavior of a variety of systems. Fisher,
Fisher and Huse, (FFH)\cite{ffh} have derived a dynamic scaling 
form for superconducting systems that should apply under a 
variety of conditions, from zero applied magnetic field to 
finite field, from ``clean" to disordered systems, as long as 
the transition is continuous. Typically, the ``FFH" scaling 
is applied to finite field $I$-$V$ curves. 

Recently the authors 
of this manuscript applied the FFH dynamic scaling to zero-field 
experimental data from two-dimensional (2D) superconducting films, 
Josephson junction arrays (JJA's), and superfuid $^4$He 
films\cite{ammirata99,pierson99} to find a value $z\simeq5.6$ for 
the dynamical critical exponent. The commonly expected and reported 
value of $z$ for these systems is $2$, which represents simple 
diffusion. In this paper, we extend our work to simulational
$I$-$V$ data on the neutral 2D Coulomb gas and find $z\simeq 5.6$.

The traditional form of the FFH scaling is
$V=I\xi^{-z} \chi_\pm(I\xi/T),$ where $\chi_{+ (-)} (x)$ 
is the scaling function for temperatures 
above (below) the transition temperature $T_{c}$ and 
$\xi$ is the vortex correlation length.  
We use a form that is slightly different:\cite{note}
\begin{equation}
I^{1+1/z}/[TV^{1/z}]=\varepsilon_\pm(I\xi/T)
\label{FFHsc}
\end{equation}
where $\varepsilon_\pm (x)\equiv x/\chi_\pm^{1/z} (x)$. 
The left-hand side of this equation does not
involve any factors of $\xi$, which tend to stretch the scaling
axes, thereby diminishing the resolution. Thus Eq.~(\ref{FFHsc}) 
allows one to better judge the scaling, as we mention below.  

\begin{figure}[b]
%h=here, t=top, b=bottom, p=separate figure page
\begin{center}\leavevmode
\includegraphics[width=0.8\linewidth]{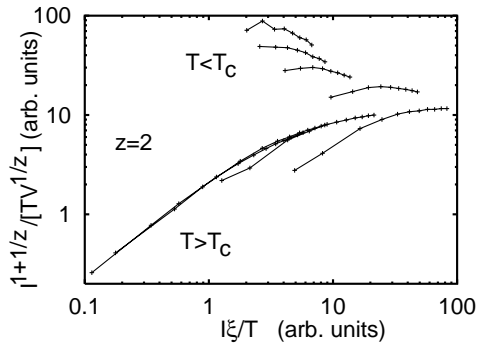}
\caption{ 
Simulational $I$-$V$ data of Lee and Teitel\protect\cite{leeteitel},
scaled according to Eq.~(\protect\ref{FFHsc}), with $z=2$.  
The scaling parameters were determined in Ref.~\cite{leeteitel}, 
using finite-size scaling.
}\label{compare2}\end{center}\end{figure}

We demonstrate the application of 
Eq.~(\ref{FFHsc}) for the numerical data of Lee and 
Teitel\cite{leeteitel}.  Using a finite-size scaling analysis, 
those authors found a dynamical exponent of $z=2$.   
The parameter $C=0.35$ entered through the FFH scaling using 
the conventional form for the Kosterlitz-Thouless correlation 
length, $\xi=\exp[C/\left\vert T -T_c\right\vert^{1/2}]$, 
while the transition temperature, $T_c=0.218$, was determined 
in equilibrium simulations\cite{leeteitel}.  The results are 
shown in Fig.~\ref{compare2}.  The collapse is not convincing, 
especially for the $T<T_{c}$ isotherms. (It is instructive to
contrast Fig.~\ref{compare2} with the Fig.~5 of 
Ref.~\cite{leeteitel}, where the scaling is presented for 
the same parameter values but in the original FFH dynamic 
scaling form. This comparison illustrates the advantage of 
using Eq.~(\ref{FFHsc}) to judge the scaling collapse.)

\begin{figure}[t]
%h=here, t=top, b=bottom, p=separate figure page
\begin{center}\leavevmode
\includegraphics[width=0.8\linewidth]{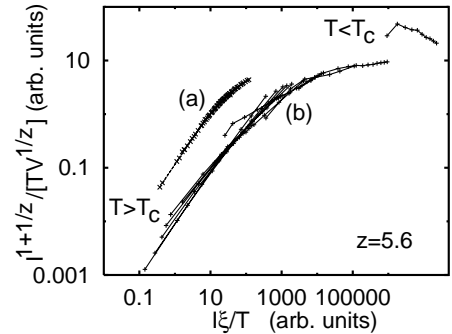}
\caption{ 
(a) The Lee-Teitel $z=5.6$ results\protect\cite{leeteitel}; and (b) the
Weber {\it et al.} $z=5.6$ scaling results\protect\cite{weber96}.
}\label{simulate}\end{center}\end{figure}

By varying the scaling parameters $T_c$, $z$, and $C$ in a 
least-squares minimization of the separation between
scaled data, 
we have obtained a more convincing collapse of the Lee-Teitel
data, as shown in Fig.~\ref{simulate}, curve (a).  According to
the dynamical scaling hypothesis of FFH, the critical isotherm,
$T=T_c$, is expected to form a perfect power-law, $V\propto I^{z+1}$, thus
appearing straight in a log-log plot of $I$ and $V$.  Generally, it is
preferable to identify $T_c$ in this way, permitting only small variations
from this estimate in an $I$-$V$ fitting
procedure\cite{ammirata99,pierson99}. However, according to the power-law
criterion, the Lee-Teitel data all appear to lie in the high temperature
range, $T>T_c$.  We therefore allow $T_c$ to remain an unrestricted
parameter, obtaining the fitting results $T_c=0.122$, $z=5.6$, and
$C=1.13$.  We performed a similar analysis on the Monte Carlo data of
Weber {\it et al.}\cite{weber96} and found an optimal collapse for
$T_c=0.131$, $z=5.6$, and $C=1.463$, as shown in Fig.~\ref{simulate},
curve (b). These values of $T_c$ and $z$ are combatible with the power-law
criterion.

The similarities between scaling results in experimental 
systems\cite{ammirata99,pierson99} and the simulations studied here
are striking.  In every case, optimization produces unexpectedly high, but
universal values of $z\simeq 5.6$. However, the results presented here
raise some important questions.  It is apparent that the critical $I$-$V$
isotherm, separating superconducting from nonsuperconducting behavior,
occurs significantly below the equilibrium ($I=0$) transition temperature.
 It is not yet clear whether this lower transition temperature is an
artifact of finite-size effects, or represents a more fundamental
phenomenon in 2D vortex dynamics.  We intend to address this issue in the
future.

\begin{ack}
SWP acknowledges the The Petroleum Research Fund, 
administered by the ACS, for their support.

\end{ack}

\end{document}